# How random is your heart beat?

Krzysztof Urbanowicz[1,2], Jan J. Żebrowski[1], Rafał Baranowski[3] & Janusz A. Hołyst[1]

[1]*Faculty of Physics, Warsaw University of Technology, Koszykowa 75, PL-00-662 Warsaw, Poland*

[2]*Max Planck Institute for the Physics of Complex Systems, Nöthnitzer Straße 38, 01187 Dresden, Germany*

[3]*Institute of Cardiology at Warszawa, Alpejska 42, 04-628 Anin, Poland*

**Abstract:**

We measure the content of random uncorrelated noise in heart rate variability using a general method of noise level estimation using a coarse grained entropy. We show that usually - except for atrial fibrillation - the level of such noise is within 5 - 15% of the variance of the data and that the variability due to the linearly correlated processes is dominant in all cases analysed but atrial fibrillation. The nonlinear deterministic content of heart rate variability remains significant and may not be ignored.

## Introduction

The standard ECG trace that our physician examines looks regular. The pattern visible in the printout of the electrocardiogram seems to repeat itself while in fact the time intervals between heart beats usually change in a complex and irregular way. This phenomenon called heart rate variability is observable when the proper time resolution is used (tens to hundreds of miliseconds of change from heart beat to heart beat). A variety of physiological factors affect human heart rate. It is now well known that the properties of heart rate variability may be an important factor in the assessment of serious cardiac conditions especially of the risk of sudden cardiac death [1]. An open question is the source of heart rate variability.

The normal heart cycle begins with the electrical activity of a specialized group of cells in the right atrium of the heart – the sino-atrial node SA - which acts as the principal pacemaker of the heart. The action potential then propagates along the atria reaching the second node of the heart - the atrio-ventricular node AV. Reacting to the potential of the AV node, the His-Purkinje system of fibers inside the ventricles delivers stimuli at different locations allowing the ventricles to contract in a concerted way. Both branches of the autonomous nervous system act on the SA and the AV node, constantly moderating the heart rate. The activity of the nervous system is a function of a number of feedback loops, of which the one controlling blood pressure (the baroreceptor system) and the one keeping the level of carbon dioxide in the blood at bay (the chemoreceptor system) play a decisive role [2]. All together the interplay of at least five nonlinear oscillatory processes affect the human blood distribution system [3,4] and so the heart rate.

Heart rate variability is measured as a time series of the time intervals between successive contractions of the ventricles of the heart (i.e. the RR intervals of the ECG recording). If the sinus node is the pacemaker responsible for the heart rhythm then such a rhythm is called sinus rhythm. It is the most common and natural rhythm.

An important aspect of the heart rate variability generation process is the interplay between sinus rhythm and the propagation of the action potentials in the form of waves originating in the SA and AV nodes within the atria and the ventricles. In certain circumstances, parts of the heart tissue may become self-oscillatory (a property called automatism) so that various forms of arrhythmia in the atria and in the ventricles occur, disrupting the normal (sinus) rhythm [2]. In particular, during atrial fibrillation, when a break up of the waves occurs within the atria, the heart rate variability is so large and complex that the rhythm is accepted to be random [5].

However, even without arrhythmia, the variability of sinus rhythm in a healthy individual is very complex (fig.1). It is now accepted that, in general, disease as well as age may result in a decrease of heart rate variability. Denervation of the heart due to cardiac infarction or heart transplant reduces heart rate variability severely. In clinical practice, standards exist for the measurement of the properties of the variability of heart rate [1] - a means of assessing the state of the heart rate control system, mainly that of the autonomous nervous system. In this context, both time domain (e.g. standard deviation of the heart rate) and frequency domain methods (power spectral analysis) are used for diagnostic purposes and the assessment of the risk of sudden cardiac death, in particular. In many cases these methods are ineffectual: the standard deviation of the heart rate for both healthy individuals and for the high risk patients may be indistinguishable (fig.2) while a large number of arrhythmia in the heart beat sequence renders frequency analysis of the sinus rhythm useless [1].

For this reason, both a search for better diagnostic tools for heart rate variability analysis is under way and the sources of the variability itself are researched. Both goals are, of course, closely related. Considerable effort has gone into methods based on the assumption that, in view of the complexity of the activity of the autonomous nervous system, at least a major part of the variability of the heart rate may treated as a noise driven process [6,7,8]. Most of these methods use fractal or multifractal scaling analysis [9]. The approach has also yielded stochastic models of heart rate variability [10,11].

Researchers using a predominantly deterministic approach [12,13,14] also claim success in the description of heart rate variability and propose various new methods for medical diagnostics. In some cases, such examples of typical deterministic structures in phase space as hyperbolic saddles [15] may be identified in heart rate variability data in the presence arrhythmia but also spiral trajectories around a saddle-focus [16] for pure sinus rhythm may be found.

We see then that the question what is the random (or noise) content in heart rate variability and how much of this phenomenon is due to the deterministic, nonlinear instability of the system is open and valid. . It is now accepted that the level of complexity of heart rate variability decreases both with age and several kinds of pathology (see e.g.[17,18,19]). In this paper, using a general method [20], we measure the content of random, uncorrelated noise in heart rate data. We analyzed 70 24-hour recordings of heart rate variability measured in patients of the Institute of Cardiology at Warszawa. We show that usually - for both cases of disease and for healthy individuals - the level of random, uncorrelated noise is within 5 - 15% of the variance of the data. The exceptions were cases of atrial fibrillation where the level of random noise was found to exceed 60 %. We also show that linearly correlated processes are dominant in heart rate variability but with the advent of disease and of the risk of cardiac arrest the nonlinearly correlated components increase. We demonstrate that removing the random noise content from the data uncovers the deterministic trajectories obscured by it. Our results show that heart rate variability has an important component due to linearly correlated processes but that the role of deterministic processes is significant.

The results we present are different from those usually presented according to the Task Force Standards [1] for the analysis of heart rate variability. Firstly, the latter are based on a time series of RR intervals with the arrhythmic beats removed by filtering. We do not filter the data to remove arrhythmia. Secondly, the standards are based on definitions of the bandwidths ULF, LF and HF to which some physiological interpretation is attributed. We analyze the degree of randomness of heart rate variability without referring to the standard bandwidth and analyze how correlated the variability is. Note that in most of the cases studied by us other than the normals, the standard procedure of analyzing only normal heart beats can not be applied because of the large incidence of arrhythmia in these cases i.e. the well established clinical techniques for heart rate variability analysis cannot be applied to most of our data.

**Methods**

To estimate the level $N$ of random noise in the analyzed data, we applied a method developed by Urbanowicz and Hołyst [20]. All data sets were analyzed by means of a 1000 data point sliding window shifted by 200 RR intervals. The method makes use of the properties and theorems of deterministic dynamical systems and chaos theory. If noise is added to the trajectory of a deterministic dynamical system (measurement noise) or if noise is present in the equation of motion (dynamical noise affecting the dynamics of the system) then the complexity measure called coarse-grained correlation entropy $K_2$ [21] increases. Knowing the analytical dependence of this entropy on the standard deviation of the uncorrelated noise $\sigma$ [20], we can estimate the noise level from the calculation of the entropy $K_2$. The method also allows to estimate the error of the standard deviation of the random noise $\sigma$.

The level of random noise in the data $N$ we define as the ratio of the variance $\sigma^2$ of the random noise to the variance of the data. We express it and all similar quantities in percent.

The noise level estimation method [20] was developed for uncorrelated noise. In the case of a highly correlated stochastic process, our analysis may underestimate the stochastic component of the variability. This is because the method is sensitive to the occurrence a strong autocorrelation in the data which results in the appearance of deterministic lines in the recurrence plots [22,23,24]. This affects the resultant $K_2$ entropy. For example, if the underlying process was a linear stochastic process such as the highly correlated ARMA [25], the noise level would be estimated at below 100%.

In order to better understand better the results of the estimation of random noise, we calculate an additional parameter $L$ - the level of linear correlations in the data that biases our noise level estimation. This parameter is calculated as the maximal absolute value of autocorrelation function for the delay $\tau \in [1,10]$ with the delay expressed in the indices of the RR intervals. Increasing the maximal value of delay beyond 10 does not change the results significantly.

The meaning of this parameter can be described as follows. Bias in noise level estimation may occur due to a deterministic term present in system dynamics which leads to linear correlations in the observed time series. In the case of the well-known stochastic equation ARMA $y_{i+1} = ay_i + \xi_i$, the level of bias can be measured by the parameter $a$. On the other hand the parameter $a$ can be calculated from the value of autocorrelation function with a delay 1: $AC(1)$.

Now let us consider a more general problem when the system possesses a periodic solution with a period $\tau$ and the dynamics is contaminated by noise. A relative deviation from the periodic behaviour can be calculated as

$$\%Noise\_in\_periodic\_signal = \frac{\left\langle (x_i - x_{i-\tau})^2 \right\rangle}{\left\langle (x_i^{shuffle} - x_{i-\tau}^{shuffle})^2 \right\rangle} = 1 - \%Periodic\_Signal = 1 - L(\tau)$$

where $\left\{ x_i^{shuffle} \right\}$ is a shuffled time series $\left\{ x_i \right\}$.

It is clear that the parameter $L(\tau)$ is the measure of periodical component and can be calculated as

$$L(\tau) = 1 - \frac{\left\langle (x_i - x_{i-\tau})^2 \right\rangle}{\left\langle \left( x_i^{shuffle} - x_{i-\tau}^{shuffle} \right)^2 \right\rangle} = 1 - \frac{\left\langle x_i^2 \right\rangle + \left\langle x_{i-\tau}^2 \right\rangle - 2\left\langle x_i x_{i-\tau} \right\rangle}{\left\langle \left( x_i^{shuffle} \right)^2 \right\rangle + \left\langle \left( x_{i-\tau}^{shuffle} \right)^2 \right\rangle} = \frac{\left\langle x_i x_{i-\tau} \right\rangle}{\left\langle \left( x_i^{shuffle} \right)^2 \right\rangle} = AC(\tau)$$

where $AC(\tau)$ is an autocorrelation function with delay τ. The appearance of the second moment in the last equation is due to the fact that we would like to express $L(\tau)$ in percent and that the variances of independent signals are additive. We see that for the case of ARMA as well as for our general periodic case discussed above the maximum of autocorrelation function can be a measure of system determinism. If we do not know the period τ a priori, or if we do not know what type of signal we have, we should check all possible values looking for the maximum of $L(\tau)$ at τ > 0. In all these cases the parameter L can serve as the value for bias of our main parameter N.

We used the noise reduction method *Local Projection with Nonlinear Constraints* (LPNC) [26] that applies a linear approximation of an attractor in the nearest neighbourhood in the Takens space [27]. The main difference between the standard method of Local Projection [27] and the LPNC are the nonlinear constraints that appear in a natural way in deterministic systems. Because the constraints are used together with the calculation of the Jacobi matrix (the latter is usual for standard Local Projection Methods), the numerical errors in the estimation of the elements of this matrix do not cause large errors in the corrections to the trajectory estimated by the LPNC algorithm. A second important feature of the LPNC method is the possibility of an automatic termination of the calculation at an optimum. The main input parameter of our method is the minimal projection dimension which is a function of the attractor dimension of the clean trajectory (here we used 4). The remaining parameters can be set to default values so that the method can be used for noisy data from an unknown source. Note that the noise reduction method may be used only for a moderate level of noise [28].

The source and executables for the methods [20,26] can be found at the site http://www.chaosandnoise.org

**The Data**

For all cases discussed below, heart rate variability data was extracted from 24-hour Holter ECG recordings analyzed with the 563 Del Mar Avionics system. All data were checked by a qualified cardiologist: normal beats were detected, artifacts were deleted and arrhythmias were recognized. The data was sampled at 128 Hz. Contrary to the usual clinical practice, the arrhythmias were not filtered out. For most of the cases studied by us other than the normals, the standard procedure of analyzing only normal heart beats can not be applied because of the large incidence of arrhythmia. Thus, the well know techniques of standard heart rate variability analysis[1] cannot be used for most of our data.

The population studied consisted of 70 24-hour Holter ECG recordings. 34 of these were recorded in healthy individuals: 28 men (mean age 37 y ± 10 y) and 6 women (mean age 34 y ± 13 y). In 7 cases (4 men and 3 women, age 63 ± 9 y) sustained atrial fibrillation was recorded during all 24 hours and in one case atrial flutter was present. Finally, 15 cases at risk of cardiac arrest (CA) were analyzed. In this group were 14 male and 1 female, 41±8 yr of age, who belonged to the highest cardiological risk group and who had experienced cardiac arrest due to ventricular fibrillation (14 patients) or asystole (1 patient). Ventricular fibrillation was a complicating factor in the coronary disease of 11 patients from this group, one person had valvular heart disease, and one had arrhythmogenic right ventricular disease. For two cases in this group, no apparent heart disease was found, except for recurrent ventricular fibrillation. Three patents had their arrhythmic event occur while wearing the Holter device; one of them died. Two other patients subsequently died suddenly, while one had repeated recurrence of ventricular arrhythmia. Following standard medical practice, each of the 15 persons from the high risk group had an age, sex, and disease status matched pair serving as the control (two of the controls – young, healthy men - were included also into the group of normals).

For noise reduction we chose two characteristic cases to demonstrate the effect of the removal of random noise from heart rate variability. The first, labelled LCH [15,16], was a post myocardial infarction patient 64 y of age, who died of ventricular fibrillation some time after the recording was made. Over 70 % of the RR intervals recorded were due to arrhythmia. The second case, CHM, was recorded in a healthy, 25 y old man with sinus rhythm and no arrhythmia.

**Random noise in heart rate variability**

We assess the level of two categories of variability within the signal: the variability of the data due to linearly correlated processes $L$ and the uncorrelated noise content $N$ – both given in per cent of the variance of the data. We analyzed the data separating it into four groups: the normals, the atrial fibrillation group, the cardiac arrest cases CA and their controls. The box plots in fig. 3 depict the four categories mentioned above.

In fig. 3a it can be seen that, for the normals, the uncorrelated noise level $N$ in the signal is relatively low (less than 10%) and that it is slightly higher for those at risk of CA and their controls. On the other hand, atrial fibrillation seems to be associated with a large level of uncorrelated noise - as expected [3]. The star symbol marks the level of uncorrelated noise for the patient with atrial flutter.

The level of heart rate variability due to linearly correlated processes $L$ for all three categories is depicted in fig.3b. The four diamond symbols mark the outliers of both the CA and of the control group. The star symbol again marks the case of atrial flutter. In keeping with the results shown in fig. 3a, when atrial fibrillation occurs this component of the variability was found to be rather low (on the average less than 20 %). On the other hand, the majority of the normals exhibit a heart rate variability with $L$ about 85 %. The controls of the CA cases have a slightly smaller level of the linearly correlated content in their heart rate variability with the group average at about 78 % and a smaller spread. The cases of a high risk of CA exhibit a somewhat smaller average but the box plot is also characteristically much wider.

Several groups have measured the complexity of heart rate variability in the past in the context of the effect of age and of pathology using a variety of nonlinear dynamics methods ranging such as multiscale entropy, recurrence diagrams and symbolic dynamics [17,18,19,29]. The approach presented differs from these in that we focus on the level of uncorrelated random noise, on the one hand, and the level of correlation in a general sense (without deciding whether the latter is of deterministic or stochastic origin), on the other.

We also assessed the potential clinical value of the methods proposed here. For a cardiologist, clinical value is in the accuracy of the separation of the group of normals from the cases of disease and in the level of the prediction of sudden death. Stepwise discrimination analysis using the SPSS system applied to all groups except atrial fibrillation (which can be immediately identified from the ECG) shows that when only the standard deviation of the heart rate and the mean heart rate were used as independent variables an overall correct classification of only 67% was obtained (75% of the normals were properly classified and only 57% of the cases of disease were correctly identified). We tried different combinations of independent variables and found that the best results were obtained using a stepwise procedure. The variables that remained in the analysis were $N$ and $L$ while standard deviation was rejected. An overall correct classification of 85 % was obtained in separating the normals from the CA cases and their controls. In this case, correctly identified were 83 % of the normals and 87% of the cases of disease. An attempt to use the same procedure to separate the study population into three categories - the normals, the cases of CA and their controls - yielded a somewhat worse result: overall accuracy 69,8% with correctly classified 82.5% of the normals , 71.5 % of the controls and 40 % of the CA cases. Note that the pathologies in the group of patients in our population were very diverse and that it is well known that no parameter or group of parameters can be universal in the risk stratification for all cardiologic pathologies.

**Heart rate variability without random noise**

Fig.4 depicts the Poincaré map of a fragment of the time series (2000 RR intervals) for the patient LCH[15], who had long lasting trigemini. before noise removal (Part a) and after (Part b). In this case the random noise component is exceptionally low although the standard deviation is large: 308 ms. The noise reduction algorithm removed 2.5% of the original signal and retained the main feature of the evolution: the RR intervals are practically periodic with only a small spread. Such a long lasting periodic behaviour as shown in fig.4 is relatively rare and it is an example of a strong interaction between the arrhythmia (over 70 % of the RR intervals in this recording) and the sinus rhytm [15,16].

Fig.5 depicts the same procedure of noise removal for the case CHM [16], a normal. In this case, the part of the signal removed by the noise reduction procedure was 9%. In effect, a well visible set of spiral trajectories may be seen. Such spiral trajectories occur in a vast majority of the normals [16]. Note that the LPNC procedure used for the removal of the random noise requires multiple recurrences to occur in phase space to be successful. Such multiple recurrence behaviour does not occur in a linearly stochastic process even if it is linearly correlated.

**Conclusions**

We measured the level of two components in human heart rate variability for 70 characteristic examples of patients and normals: the level of uncorrelated noise (the random component) and the level of the component due to linearly correlated variability.

The random component in atrial fibrillation was large as expected. Atrial flutter reduced the level of random noise in heart rate variability as compared to atrial fibrillation. The random component of heart rate variability for the other three categories studied here (the normals, the cases of CA and their controls ) was found to be in a surprising narrow range: the average not exceeding 10% for the normals while it was about 15 % for the controls of the high risk group and for the high risk group itself.

It is well known that, generally speaking, with pathology and aging – heart rate variability decreases. But how this occurs is not completely understood. For example, the high frequency component of heart rate variability usually interpreted as  a sign of vagal activity is not necessarily equivalent to random (or even colored) noise. Here, we measure the level of random (uncorrelated) noise. If we define the level of pathology as equivalent to the level of the risk of CA, we show consistently how an increase in the level of pathology affects the randomness of heart rate variability (fig.3) and that correlated behavior increases with the level of pathology. This result is partially independent of what has been found by standard techniques in heart rate variability analysis such as the measurement of the LF and the HF components of power spectral density. On the other hand, this result has bearing on a) the way we analyze data i.e. the choice of methods and b) how we should model the system.

The results indicate that - except for atrial fibrillation – heart rate variability is principally due to linearly correlated processes. Thus, it may be due either to deterministic processes with strong autocorrelations or due to stochastic processes - the methods which we use are unable to distinguish the source of this component. The proportions of the components differ from group to group. Thus, in the case of the normals the dominant component in heart rate variability is due to linearly correlated processes (average 85 % in the group studied). With disease and an increase of the risk of CA, this component diminishes dropping below 80 % for the controls and to slightly above 70 % for the high risk group. We noted also a much wider spread of our parameter $L$ - characterizing the level of linear correlations - in the CA group. Our results form an interesting image of how heart rate variability changes with increasing pathology and risk as these two attributes are arranged in fig.3.

The results we present are different than those usually presented according to the Task Force Standards [1] for the analysis of heart rate variability. Firstly, the latter are based on a time series of RR intervals with the arrhythmic beats removed by filtering. We do not filter the heart rate data at all, least of all to remove arrhythmia. Secondly, the Standards are based on the definitions of the bandwidths ULF, LF and HF to which some physiological interpretation is attributed. We analyze the degree of randomness of heart rate variability without referring to the standard bandwidth and analyze how correlated this variability is. This is a separate, independent approach and may be difficult to map onto the standard approach which is based on the linear signal analysis methods of 10 years ago.

Removing the random component from the RR interval time series and plotting the data as a Poincaré map yielded spiral trajectories and periodic orbits - typical nonlinear dynamics behaviour. Note that the LNPC noise reduction method [26] removes noise due to random processes and is based on the analysis of recurrence – a deterministic property of a dynamical system. Thus, together with the above described assessment of the level of random noise and of the level of linear correlations in heart rate variability, this result demonstrates that deterministic processes are an important factor in the formation of heart rate variability.

When stepwise discrimination analysis was performed to separate the normals from the cases with a high risk of CA and their controls, the parameters that represent simple data such as the mean RR interval (i.e. the mean heart rate) and standard deviation of the heart rate (the simplest parameter which describes heart rate variability) were not adequate to the task and were rejected by the algorithm. Although the quantities calculated in this work cannot be used to assess the level of the risk of cardiac arrest, discriminant analysis shows that, using the two parameters $N$ and $L$ introduced by us, an 85 % overall accuracy may be achieved in distinguishing the normals from the group composed of the CA patients and their sex, age and disease controls. Besides a potential use in population studies, this result indicates a relation exists between disease, the level of randomness and the kind of correlations that occur in heart rate variability.

A clear consensus on the deterministic or stochastic source of heart rate variability is still lacking. We thus endeavored to assess the level of random, uncorrelated noise and found it surprisingly low while the level of correlated processes changes with the level of pathology. Together with our results on noise reduction, the results indicate a significant content of deterministic processes in heart rate variability.

Acknowledgements

K.U. would like to thank Holger Kantz for fruitful discussions. The work was supported by the COST Action P10- Physics of Risk, by Polish Ministry of Science and Higher Education, Grant No. 134/E-365/SPB/COST/KN/DWM 105/2005-2007, and by a special Grant of Warsaw University of Technology.

Figure Captions

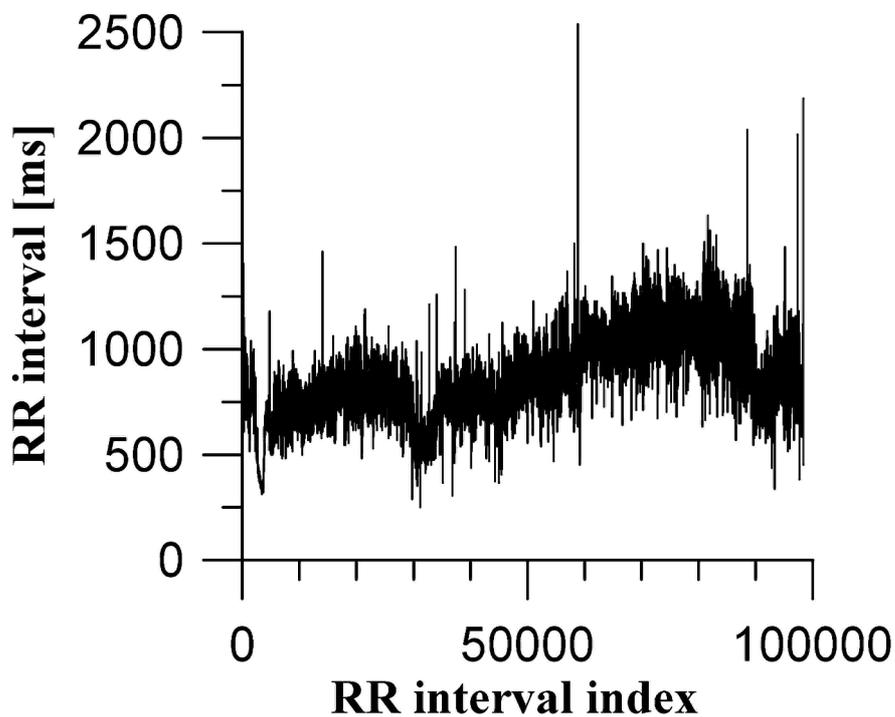

Fig.1 24-hour time series of the time intervals between heart beats measured as the RR intervals of the ECG recording for the normal CHM. The heart rate variability was decrease at around index 500 due to a 10 minute exercise stress test. The pauses exceeding 1.5 s are normal.

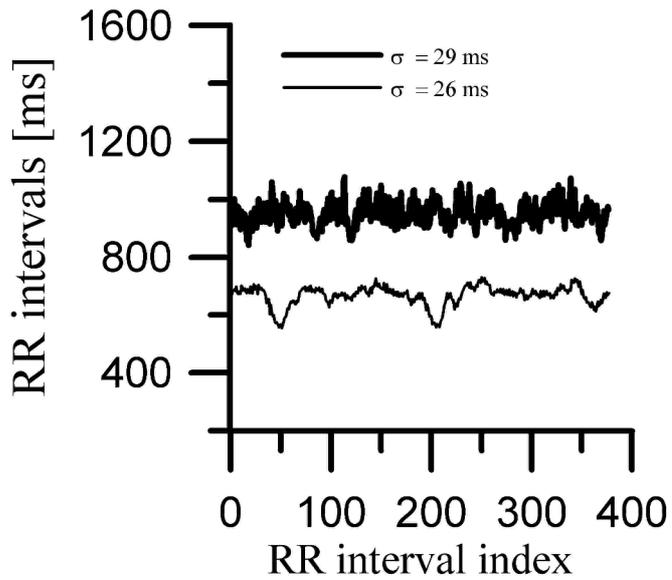

Fig.2 Comparison of two examples of heart rate variability with indistinguishable standard deviation. The thick curve was measured in a normal individual while the thin one – in a case of a high risk of cardiac arrest.

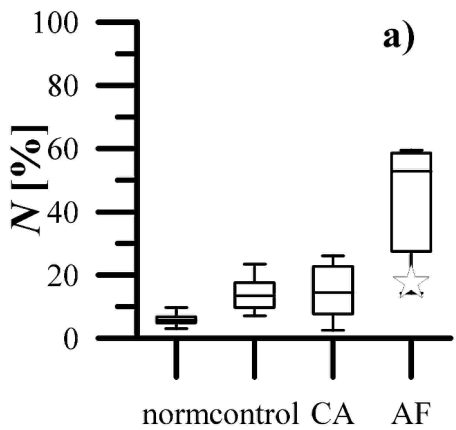

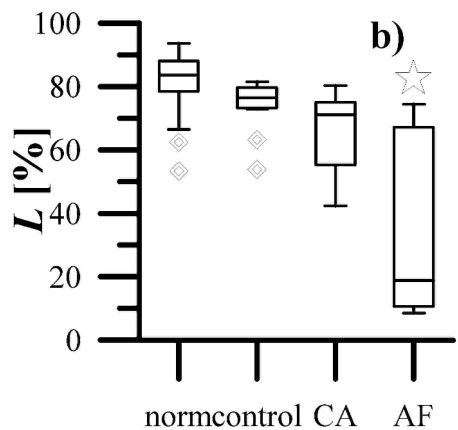

Fig.3 Box plots for the patient categories: normals, high risk of cardiac arrest, their controls and the cases of atrial fibrillation. The symbol marks the case of atrial flutter and the □ symbols – the outliers of the boxplot in the CA and the control groups. Part a: heart rate variability component due to random noise, Part b: that due to linearly correlated processes.

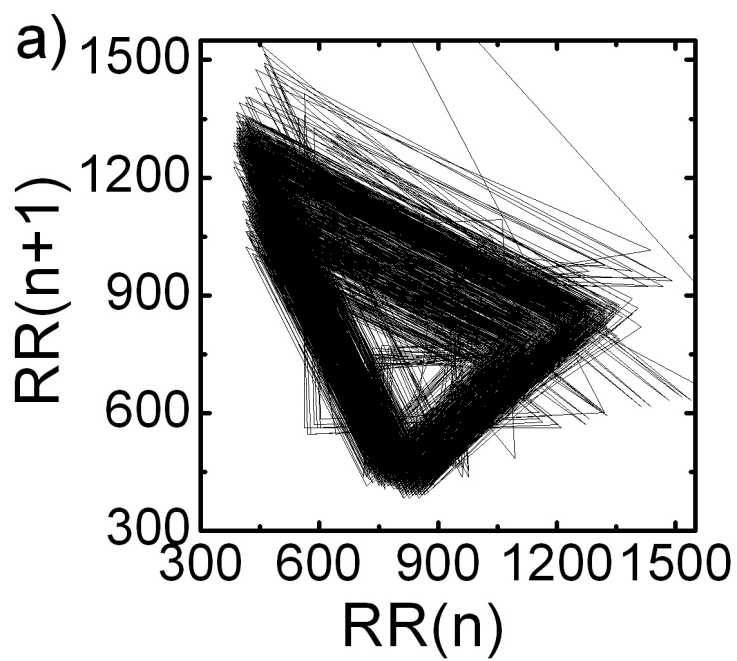

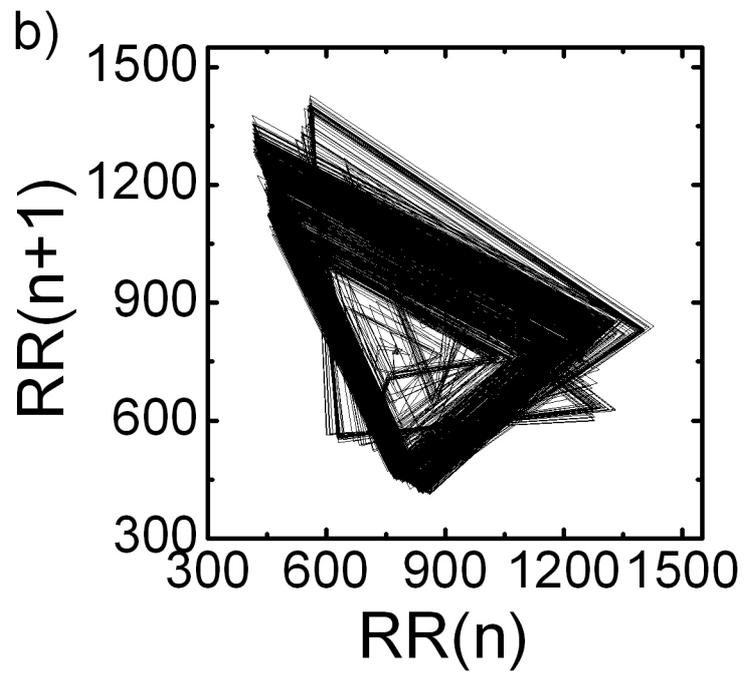

Fig.4 Effect of the removal of random noise from the RR interval time series for the case of a high risk of cardiac arrest LCH. Part a: raw data. Part b: after the application of the LNPC algorithm. The time series was 3000 data points long.

a)

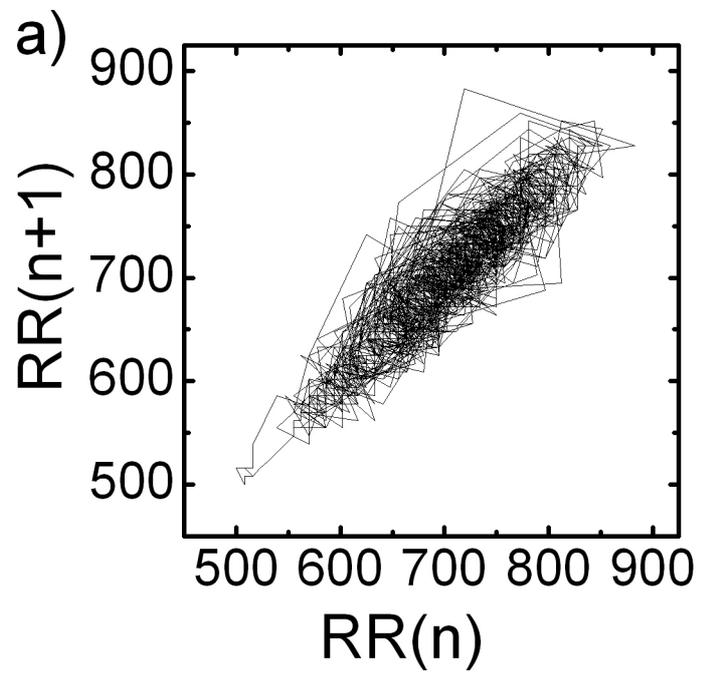

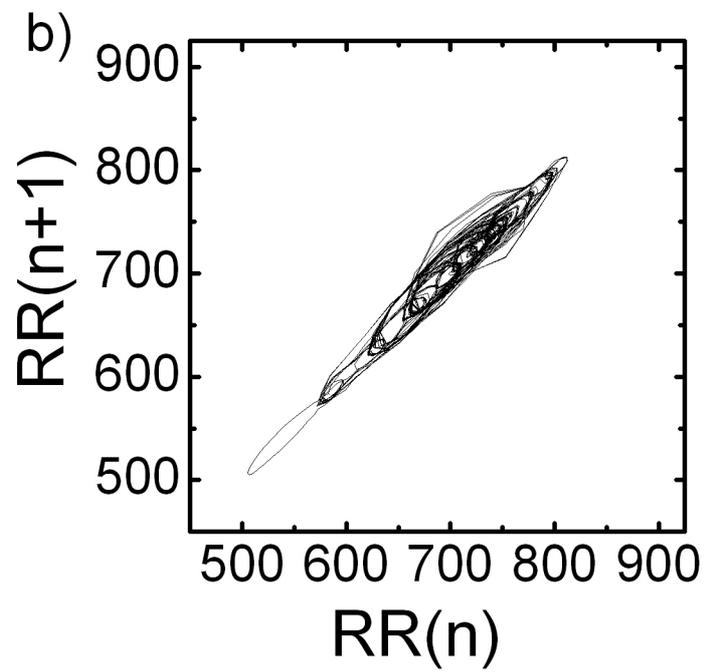

Fig. 5 The same as in fig.5 but for the normal CHM. The spiral trajectory visible in Part

b is very typical for healthy individuals.